\documentclass[12pt,preprint]{aastex}

\shorttitle{Initial-Final Mass Relation from Wide Binaries}
\shortauthors{Zhao et al.}

\begin{document}


\title{The Initial-Final Mass Relation among White Dwarfs in Wide Binaries}


\author{J. K. Zhao\altaffilmark{1,2}, T. D. Oswalt\altaffilmark{1}, L. A. Willson\altaffilmark{3}, Q. Wang\altaffilmark{3}, G. Zhao\altaffilmark{2}}


\altaffiltext{1}{Florida Institute of Technology, Melbourne, USA, 32901; jzhao@fit.edu; toswalt@fit.edu}
\altaffiltext{2}{Key Laboratory of Optical Astronomy, National Astronomical Observatories, Chinese Academy of Sciences, Beijing, 100012, China; gzhao@bao.ac.cn}
\altaffiltext{3}{Department of Physics and Astronomy, Iowa State University, Ames, IA 50010, USA; lwillson@iastate.edu; wqinisu@iastate.edu}


\begin{abstract}
We present the initial-final mass relation derived from 10 white dwarfs in wide binaries that consist of a main sequence star and a white dwarf. The temperature and gravity of each white dwarf was measured by fitting theoretical model atmospheres to the observed spectrum using a $\chi^{2}$ fitting algorithm. The cooling time and mass was obtained using theoretical cooling tracks. The total age of each binary was estimated from the chromospheric activity of its main sequence component to an uncertainty of about 0.17 dex in log \textit{t} The difference between the total age and white dwarf cooling time is taken as the main sequence lifetime of each white dwarf. The initial mass of each white dwarf was then determined using stellar evolution tracks with a corresponding metallicity derived from spectra of their main sequence companions, thus yielding the initial-final mass relation. Most of the initial masses of the white dwarf components are between 1 - 2 M$_{\odot}$. Our results suggest a correlation between the metallicity of a white dwarf's progenitor and the amount of  post-main-sequence  mass loss it experiences - at least among progenitors with masses in the range of 1 - 2 M$_{\odot}$.  A comparison of our observations to theoretical models suggests that low mass stars preferentially lose mass on the red giant branch.

\end{abstract}



\section{Introduction}
Over 90 percent of all stars shed at least half their mass as they evolve towards their final state - a white dwarf (WD). The initial-final mass relation (IFMR) represents a mapping between the mass of a WD remnant and the mass of its hydrogen-burning main-sequence (MS) progenitor. It also characterizes the amount of material stars with primordial masses M $\lesssim$ 8 M$_{\odot}$ recycle to the interstellar medium. Thus, it is of paramount importance to understanding the chemical enrichment and the efficiency of star formation in galaxies. This relation is also a key constraint on stellar evolution theory.

One of the first attempts to empirically determine the IFMR was undertaken by Weidemann (1977). Reimers $\&$ Koester (1982) and Koester $\&$ Reimers (1996) presented observations of WDs in the open cluster NGC 2516 and obtained an IFMR using these WDs.  Weidemann (2000; W00 hereafter) updated the IFMR by incorporating new theoretical and observational data. Claver et al. (2001) observed six WDs in the Praesepe open cluster and determined a monotonic IFMR. Williams et al. (2004; 2009) presented an empirical determination of the IFMR based on a spectroscopic analysis of massive white dwarfs in NGC 2168 (M35). They showed that the resultant white dwarf mass increases monotonically with progenitor mass for masses greater than 4 M$_{\odot}$. Ferrario et al. (2005) re-evaluated the ensemble of data that has been used to determine the IFMR and characterized a mean IFMR about which there is an intrinsic scatter. They showed that a linear IFMR predicts a mass distribution in reasonable agreement with the Palomar-Green survey.  Kalirai et al. (2005) determined the IFMR from observations of very faint WDs in the rich open cluster NGC2099 (M37). They found stars with initial masses between 2.8 and 3.4 M$_{\odot}$ lose 70 - 75\% of their mass during post-MS evolution. Dobbie et al. (2006) also constructed a new IFMR based on 11 WDs in the Praesepe open cluster.  Rubin et al. (2008) constructed an IFMR  based on 19 spectroscopically identified WDs in NGC 1039 (M34).  Catal\'{a}n et al. (2008a) studied the IFMR using six WDs in wide binaries and suggested the IFMR may not be a single-valued function. Kalirai et al. (2008) presented constraints on the low-mass end of the IFMR using older open clusters NGC7789 (\textit{t} = 1.4 Gyr ), NGC 6819 (\textit{t} = 2.5 Gyr) and NGC6791 (\textit{t} = 8.6 Gyr). Later, Kalirai et al. (2009) extended the IFMR to lower masses using the globular cluster M4. Since most nearby clusters are relatively young, the initial masses of those WDs tend to be high.

In this paper we investigate the IFMR using wide ``fragile" binary systems containing a WD with a MS companion. The systems have relatively large orbital separations ($<a>$ $\sim$ 10$^{3}$ AU; Oswalt et al. 1993; Silvestri et al. 2001, 2002, 2005). Thus, one can safely assume that each component has evolved independently, unaffected by mass exchange
or tidal coupling that complicate the evolution of closer pairs. Components of a given binary are
coeval (Greenstein 1986). Essentially, each may be regarded as an open cluster with only two components.  Although it is difficult to obtain ages for wide binaries as accurate as ages for open clusters, they tend to be nearer, brighter and are far more numerous than nearby clusters. Moreover, they span a much more continuous range in age. Catal\'{a}n et al. (2008a; C08a hereafter) used six wide binaries to investigate the IFMR; three of the WDs  had low initial mass ($<$ 2M$_{\odot}$). Our sample contains additional WDs at the low initial mass limit.

Previous research has indicated a large scatter in the empirical IFMR. What is the source of this scatter? Kalirai et al. (2005) found some weak evidence of a metallicity dependence in the IFMR. Kalirai et al. (2007) found evidence for enhanced mass loss at extremely high metallicities by studying the white-dwarf mass distribution in
the supersolar-metallicity star cluster NGC 6791 ([Fe/H] = +0.4). Kalirai et al. (2009) found a relatively flat relation between mass loss and metallicity ([Fe/H] between -1.1 to solar metallicity) by extending these studies to WDs in the globular cluster M4.  However, a clear relation between the metallicity and scatter in the IFMR has not been demonstrated (cf. Williams 2007; Catal\'{a}n et al. 2008b, C08b hereafter). We investigated whether there is a metallicity dependence on the IFMR  in our wide binary sample using the spectra of their MS companions.

Section 2 provides an overview of the observations and
reductions for our sample. The astrophysical properties of the WDs are discussed in section 3. The MS companions are discussed
in section 4. In section 5, we present and discuss our IFMR. Section 6 compares our observations to theoretical models of post-MS mass loss. A discussion of the implications of our findings is given in section 7.

\section{Observations and Data Reduction}

 Most of the MS+WD binaries chosen for this study are from the Luyten (1979) and Giclas, Burnham $\&$ Thomas (1971) proper motion catalogs chosen by Oswalt, Hintzen $\&$ Luyten (1988). A key impetus for using such pairs for determining the IFMR is that the total lifetime of each pair is approximately the age derived from measurements of the MS component. In addition, the total age of a pair is approximately the sum of each WD component's cooling time and the MS lifetime of its progenitor.

 Table 1 gives our observed data for 11 wide binaries. Columns 1-3 list the name of each binary, its right ascension, and declination (coordinates are the original 1950 catalog epoch). Columns 4-5 present the ID and spectral type of each WD component. Columns 6-7 list the spectroscopic observation date and site.


\subsection{Spectroscopic Observations}
Spectroscopic observations were made at two observatories.  In the southern hemisphere, observations were conducted at Cerro Tololo Inter-American Observatory (CTIO) using the Blanco 4-meter telescope.  Northern hemisphere observations were conducted at Kitt Peak National Observatory (KPNO) using the Mayall 4-meter telescope.

At CTIO, the Ritchey-Chreti$\acute{e}$n (RC) Cassegrain spectrograph on the 4-meter Blanco telescope was used on two separate observing runs (February 2004 and February 2005) to obtain optical spectroscopy of wide pairs, as well as standard flux calibration stars. During these two observing runs the KPGL1 grating was used to obtain a scale of 0.95 $\rm{\AA}$/pixel (R $\sim$ 2000).  A Loral 3K CCD (L3K) was used with the RC spectrograph.
It is a thinned 3K$\times$1K CCD with 15 $\mu$m pixels.  A spectral range of approximately 3800 - 6700 $\rm{\AA}$ was achieved.

At KPNO, the RC spectrograph with the BL450 grating set for the 2nd order to yield a resolution of 0.70 $\rm{\AA}$/pixel (R $\sim$ 2000) was used to obtain optical spectra during November 2006 and July 2010 with the Mayall 4-meter telescope. The 2K$\times$2K T2KB CCD camera with 24 $\mu$m pixels was used to image the spectra.  An 8-mm CuSO$\rm_{4}$ order-blocking filter was added to decrease 1st-order overlap at the blue end of the spectrum. A spectral range of approximately 3800  - 5000 $\rm{\AA}$ was achieved.
\subsection{Data Reduction}
The data were reduced with standard IRAF\footnote[1]{IRAF is distributed by the National Optical Astronomy observatories, which are operated by the Association of Universitites for Research
in Astronomy, Inc., under cooperative agreement with the National
Science Foundation (http://iraf.noao.edu).} reduction procedures.
In all cases, program objects were reduced with calibration
data (bias, flat, arc, flux standard) taken on the same night. Data
were bias-subtracted and flat-fielded, and one-dimensional spectra
were extracted using the standard aperture extraction method. A
wavelength scale was determined for each target spectrum (including
stellar flux standards) using HeNeAr arc lamp
calibrations. Flux standard stars were used to place the spectra
on a calibrated flux scale. We emphasize that the final flux calibrations
for the targets provide only relative fluxes, as most nights were
not spectrophotometric.

The radial velocity of each MS star was determined by cross-correlation between the observed spectra
and a set of MS template spectra. The F, G and K template spectra were generated from a theoretical atmosphere grid (Castelli $\&$ Kurucz 2003). The dM template spectra were compiled using observed
M dwarf spectra from the Sloan Digital Sky Survey (SDSS)\footnote[2]{http://www.astro.washington.edu/slh/templates}. The wavelength ranged from roughly 3900 - 9200 $\rm\AA$ (see Bochanski et al. 2007). Our typical internal measurement uncertainties in radial velocity
were $\sigma\rm_{v_{r}}$ = $\pm$ 4.6 km s$^{-1}$.




\section{Analysis of White dwarfs }
\subsection{Determination of $T$$\rm_{eff}$
and log $g$}
Our sample included nine DA WDs and two DB WDs. For DA WDs the $T$$\rm_{eff}$ and log $g$ were derived via simultaneous
fitting of the H$\beta$ to H8 Balmer line profiles using the procedure outlined by Bergeron
et al. (1992). The line profiles in both observed spectra and model spectra were
first normalized using two points at the continuum level on either
side of each absorption line. Thus, the fit should not be
affected by the flux calibration. Model atmospheres used for this fitting were derived
from model grids provided by Koester (2010). Details of the input physics and
methods can be found in that reference. Fitting of the line profiles
was carried out using the IDL package MPFIT (Markwardt 2008), which is based on
$\chi^{2}$ minimization using the Levenberg-Marquardt method. This package
can be downloaded from the project website\footnote[3]{http://purl.com/net/mpfit}.
Errors in the $T$$\rm_{eff}$ and log $g$ were calculated by
stepping the parameter in question away from their optimum values
and redetermining minimum $\chi^{2}$ until the difference between this
and the true minimum $\chi^{2}$ corresponded to 1$\sigma$ for a given number of
free model parameters. Our spectra of the DB WDs were not of high enough signal-to-noise ratio (S/N) to do line profile fitting on such weak lines. For these DB WDs we adopted the $T$$\rm_{eff}$ and log $g$
from Voss et al. (2007).

Fig. 1 shows fits of the observed Balmer lines of the nine DA WDs. The derived $T$$\rm_{eff}$, log $g$ and uncertainties  are shown in columns 2-3 of Table 2. Estimated $T$$\rm_{eff}$ and log $g$ for nine DAs were also available in the literature, allowing the comparisons shown in Fig. 2. For most of the WDs, the difference between our $T$$\rm_{eff}$ value and the literature value is smaller than 1000 K and the log $g$ difference is smaller than 0.1 dex. For WD1544-374, our log $g$ is consistent with that of C08a and Kawka et al. (2007) but differs by 0.2 dex with that of Vauclair et al. (1997). For WD2253-08, the log $g$ values in the literature differ substantially, however, our value is consistent with that of C08a. In Fig. 2, the dotted line represents the unit slope relation. In general, our results are consistent with those in the literature.

Reid (1996) reported that the pair BD+44 1847/G116-16 is a non-physical pair because the gravitational redshift  K$\rm_{RS}$ = 113.9 km s$^{-1}$  corresponds to 1.1 M$_{\odot}$ which is  inconsistent with 0.75 M$_{\odot}$ from line profile fitting by Begeron et al. (1995). We determined a mass of 0.77 M$_{\odot}$ that is almost the same as Begeron et al. (1995). Although our spectral resolution is only $\sim$ 1.8 $\rm\AA$, we attempted a rough redshift measurement of this pair. The doppler shift of the WD component measured from H$\beta$ was about 120 $\pm$ 10 km s$^{-1}$ which is consistent with that of Reid (1996). The parallaxes of the components from Simbad Astronomical Database (Genova
2006) for this pair are 34.6 mas and 19.36 mas for the primary and secondary, respectively. Hence, we also conclude this pair is non-physical. It was therefore eliminated from our determination of the IFMR.

\subsection{Determination of WD cooling times and mass}
From our $T$$\rm{_{eff}}$ and log $g$ estimation for each star, its current mass (M$\rm_{f}$) and cooling time (t$\rm_{cool}$) were estimated from Bergeron's cooling sequences\footnote[4]{The cooling sequences can be downloaded from the website: http://www.astro.umontreal.ca/~bergeron/CoolingModels/.}. For the pure hydrogen model atmospheres hotter than $T$$\rm{_{eff}}$ = 30,000 K, we used the carbon-core cooling models of Wood (1995), with thick hydrogen layers of q$\rm_{H}$ = M$\rm_{H}$/M$_{*}$ = 10$^{-4}$. For $T$$\rm{_{eff}}$  cooler than 30, 000 K we used cooling models similar to those described in Fontaine et al. (2001) but with carbon-oxygen cores and q$\rm_{H}$ = 10$^{-4}$ (see Bergeron, Leggett $\&$ Ruiz 2001). For the pure helium atmospheres we used similar models but with q$\rm_{H}$ = 10$^{-10}$. M$\rm_{WD}$ and t$\rm_{cool}$ were then calculated by spline interpolation based on the $T$$\rm{_{eff}}$ and log $g$. The final M$\rm_{WD}$ and t$\rm_{cool}$ values are shown in columns 4-5 of Table 2.  The uncertainties in mass are estimated based on the uncertainties in $T$$\rm_{eff}$ and log $g$ when interpolating in the cooling sequences.
The average mass of our 11 WDs is 0.628 M$_{\odot}$. The average mass of only the 9 DA WDs is 0.635 M$_{\odot}$, which is consistent with Bergeron et al. (1995; 0.626 M$_{\odot}$).

\section{Analysis of the Main Sequence Stars}
The components of each wide binary share the same original properties such as age, chemical composition and dynamics (Greenstein 1986). Thus, the MS components of wide binaries provide valuable information about the progenitors of their WD companions.
\subsection{The age determination }
Age is one of the most difficult to
determine properties of any star. Ages for single lower MS stars derived from isochrone fitting are especially uncertain because of
the narrow vertical dispersion of isochrones within the MS in an
H-R diagram. Small uncertainties in luminosity and metallicity
propagate into large uncertainties in age. Lachaume et al. (1999) investigated the precision of age determinations  for a sample of nearby MS stars of spectral types B9-K9. They found the isochrone method is best for hot stars, while stellar rotation (Barnes 2007) is best for cool stars. The MS components of our wide binaries are F-K stars. Most of our wide binaries are older than 1 Gyr, so measurement of their rotation period would require a large investment of observing time and very precise photometry.

For decades, chromospheric activity (CA) has been known to inversely correlate with stellar age (Skumanich 1972). Early work by Wilson (1963; 1968) and Vaughan $\&$ Preston (1980) established CaII H$\&$K emission as a useful marker of CA in lower MS stars.  The most commonly used measurement index of CaII H$\&$K emission is $R\arcmin\rm{_{HK}}$, defined as the ratio of the emission from the chromosphere in the cores of the CaII H$\&$K lines to the total bolometric emission of the star. Soderblom et al. (1991), Lachaume et al. (1999) and Mamajek $\&$ Hillenbrand (2008) presented detailed investigations of the relation between $R\arcmin\rm{_{HK}}$ and stellar age.

 Soderblom et al. (2010; Fig. 8) compared ages of G dwarfs derived from isochrone placement and from HK activity index. It is clear that the isochrone ages were larger than activity ages. For some stars, isochrone ages were inconsistent with the age of Galactic disk. Most stars in our sample are late main sequence G stars and K stars, for which the isochrone method does not work well, as discussed above.

Following Hall et al. (2007), for each star we computed the flux ratio $S$$\rm_{HK}$:
\begin{eqnarray}
S\rm_{HK}&\equiv&\rm{\alpha\frac{H+K}{R+V}}
\end{eqnarray}
where H and K are the fluxes measured in 2 $\rm{\AA}$ rectangular windows centered on the line cores of CaII H$\&$K; R and V are the fluxes measured in 20 $\rm{\AA}$ rectangular `pseudocontinuum' windows on either side. These
bands are similar to those used in the Mount Wilson chromospheric activity survey program (Baliunas et al. 1995), except that the bands centered on Ca II H$\&$K are wider (2 $\rm \AA$) than those used at Mount Wilson (1 $\rm \AA$ )
because of our instrumental resolution. Here $\alpha$ is 10, indicating the psendocontinuum windows are 10 times wider than the H$\&$K windows in wavelength coverage. Fig. 3 presents our H$\&$K measurements for these 10 MS stars.

In order to derive a transformation relation between our instrumental systems and the Mount Wilson system we selected six stars which have stable CA from Baliunas et al. (1995) as `standard CA stars'. Since $S\rm_{MW}$ values for four MS stars in our CTIO sample were found in the literature, we only observed these `CA standard stars' with KPNO telescope.  Table 3 provides data for these standard stars. Column 1 lists the stars' name; columns 2 - 3 give  our $S\rm_{HK}$ and the published $S\rm_{MW}$ respectively.  Fig. 4 shows the correlation between $S\rm_{HK}$ and published $S\rm_{MW}$; Equation 2 is the transformation equation obtained by a least squares fit. The scatter is $\sigma$$_{S\rm_{MW}}$ = 0.017.

\begin{eqnarray}
S\rm_{MW}&=-0.069+1.308S\rm_{HK}
\end{eqnarray}

The range of $S\rm_{HK}$ for which Equation 2 applies is limited by the range of $S\rm_{HK}$ in `standard CA stars', which is about 0.15 $<$ $S\rm_{HK}$ $<$ 0.55. With Equation 2, the $S\rm_{HK}$ value we measured at KPNO can be directly transformed to $S\rm_{MW}$.  For wide binary stars observed at CTIO  we adopted the $S\rm_{MW}$  values from the literature wherever possible. If more than one value was available in the literature, reference was given to those derived from higher resolution spectra. This is based on the conclusion by Jenkins et al. (2011) that CA measures from low resolution spectra significantly increase the rms scatter when calibrating onto a
common system such as the Mt. Wilson system. The $S\rm_{MW}$ values of CD-59 1275 and CD-38 10983 were obtained from Henry et al. (1996); 40 Eri A came from Baliunas et al. (1995) and  CD-37 10500 was found in Jenkins et al. (2011).  Only two MS stars observed at CTIO did not have $S\rm_{MW}$ values in the literature. For these, the $S\rm_{HK}$ (CTIO) were first calibrated into $S\rm_{HK}$ (KPNO) using the empirical relation  between the two instrument (Zhao et al. 2011a):    $S\rm_{HK}$ (CTIO) = $S\rm_{HK}$ (KPNO) + 0.095. Then, the  $S\rm_{MW}$ of these two stars were computed using Equation 2. For the wide binary MS stars observed at KPNO, the $S\rm_{MW}$ were directly calculated from Equation 2.

The $S\rm_{MW}$ index includes both photospheric and chromospheric
contributions. The photospheric flux
can be removed approximately using the procedure of
Noyes et al. (1984), who derived a quantity $R\rm_{HK}$ $\propto$ $F\rm_{HK}$/$\sigma$ $T^{4}\rm_{eff}$
, where $F\rm_{HK}$ is the flux per square centimeter in the H
and K band passes. The quantity $R\rm_{HK}$ can be derived from
$S\rm_{MW}$ by modeling the variation in the continuum fluxes as a function of effective temperature
(using B-V as a proxy). $R\rm_{HK}$ must then be further corrected
by subtracting the photospheric contribution in the cores
of the H and K lines. The logarithm of the final quantity $R\arcmin\rm_{HK}$ is a useful measure of the chromospheric emission, essentially independent of the effective temperature.

We computed $R\arcmin\rm_{HK}$ for the six stars observed at CTIO and the five stars observed at KPNO. Their colors (B-V) meet the condition of Noyes et al. (1984; B-V $<$ 1.1). Both the estimates of $S\rm_{MW}$ and log $R\arcmin\rm_{HK}$ are tabulated for our program stars in Table 4.

CA vs. age relations were published by Soderblom et al. (1991); Donahue (1998); Lachaume et al. (1999) and Mamajek $\&$ Hillenbrand (2008). Rocha-Pinto $\&$ Maciel (1998) investigated the metallicity dependence of the Soderblom et al. (1991) relation. Because we intended to explore the effects of metallicities within our sample, we used their relation to estimate our ages.


\begin{eqnarray}
\rm{log}~t&=&(-1.50\pm0.003)\rm{log}~R\arcmin\rm_{HK}+(2.25\pm0.12)
\end{eqnarray}

\begin{eqnarray}
\Delta\rm{log}~t&=&\rm{-0.193-1.382[Fe/H]-0.213[Fe/H]^{2}+0.270[Fe/H]^{3}}
\end{eqnarray}

The total age  estimates and uncertainties for our program stars derived from Equations 3 - 4 are listed in column 6 of Table 4.  The age uncertainties originate mainly from the $S$$\rm_{HK}$ uncertainties as they affect Equation 3. The $S$$\rm_{HK}$ uncertainties are determined from the average of the standard deviations of measurements for stars with more than two observations (about $\pm$4.6\%). The total internal uncertainty of our age determinations is about $\pm$0.17 dex. Independent age determinations for a couple of pairs were found in
the literature. These pairs are listed in Table 5. Column 1 gives
their identifications. Columns 2, 3 and 4 list the ages included
in Holmberg et al. (2009), Valenti $\&$ Fischer (2005) and Barnes (2007),
respectively. Our ages are younger than
isochrone fitting ages in all four cases. For 40 Eri A, the isochrone
fitting age is unreasonably large, while our age for this star is
close to the rotation age 4.75 $\pm$ 0.75 Gyr from Barnes
(2007). For CD-38 10983, our age is also consistent with the rotation age, but younger than the isochrone age, which has a large uncertainty. For CD-59 1275, our age is close to the isochrone age. For CD-37 10500, the error bar is too large for the isochrone age to be useful. Thus, even in this small sample it can be seen how difficult it is to get consistent isochrone ages for
late G and K dwarfs. Small uncertainties in luminosity and metallicity
cause large uncertainties in age.

Ages from isochrone fitting of clusters with well defined turnoffs are unquestionably more accurate than those of single field stars.  To estimate the accuracy of our CA age estimates, we selected 75 member stars in M67 having $R\arcmin\rm_{HK}$ and B-V data from Mamajek $\&$ Hillenbrand (2008).  All the original HK observations of these objects are from Giampapa et al. (2006).  The CA age of each member star was estimated with the above formula.  Fig. 5 is the resulting CA age distribution of this sample.  The dotted line is a Gaussian fit.  The average age of those member stars is 3.28 Gyr with 1 $\sigma$ 0.95 Gyr (1-sigma uncertainty of 29\%).  The difference between the average CA age for this cluster and its turnoff age (4.0 Gyr) is about 0.73 Gyr.  This difference, about 18\%, is well within the formal uncertainties of our wide binaries' age determinations.

All are M67 stars discussed above are solar-type dwarfs, which are likely to have activity cycles, perhaps similar to the Sun's 11-22 year period.  The stars with very high HK values could have been observed at maximum activity during their cycle (perhaps a few are close interacting binaries).  The stars with very low HK values could be in a Maunder minimum state.  Most likely these are the main sources of scatter in CA ages derived from coeval populations of stars like clusters.  We did not attempt to quantify these factors here because each of our objects has so far only been observed a small number of times.  This is almost certainly the dominant source of scatter in CA ages.

	As a final reality check, we randomly drew 10 stars from the data shown in Fig. 5 and computed the age and 1 $\sigma$ estimates.  The age of M67 derived from a 10-star sample is 3.3 Gyr $\pm$ 0.9, which represents a 28\% uncertainty-very comparable to the 30\% claimed by Soderblom (2010) and others for CA age determinations.  In our 10 wide binary sample the formal age uncertainty is 0.17 dex (48\%; computed from uncertainties listed Table 1).  We do not propose that the age of any single star has been determined to a precision any better than this.  These uncertainties have been fully propagated into the error bars shown in the plots shown in Figs. 6 and 7 below.

The difference between the total age of a MS star and the cooling age of its WD companion provides an estimate of the latter's progenitor MS lifetime (given in column 6 of Table 2).



\subsection{The metallicity measurements}
The metallicity of each MS star was measured by comparing the observed spectrum to a set of template spectra. Initially, a library of low resolution theoretical spectra was generated using the SYNTH program based on Kurucz¡¯s New Opacity Distribution Function atmospheric models (Castelli $\&$ Kurucz 2003). The details can be found in Zhao et al. (2011a, b). Fig. 3 of Zhao et al. (2011a) presents an example of how we determined stellar  metallicity and its uncertainty in 40 Eri A. The final [Fe/H] values are given in column 7 of Table 2. In Sec. 6, the metallicity is indicated by Z, where [Fe/H] = 0.0 is equivalent to Z = 0.019.

\section{Initial-final mass relation}
Once the MS lifetimes of the WDs and the metallicities of their companions were obtained, the WD initial masses were estimated from evolution models (Girardi et al. 2000) by interpolation among the tables for Z = 0.030, 0.019, 0.008 and 0.004. The initial (M$\rm_{i}$) and final masses (M$\rm_{f}$) obtained are listed in columns 8 and 4 of Table 2, respectively.

Fig. 6 displays the resulting empirical IFMR obtained for 10 WDs in our sample (filled circles). Diamonds represent the WDs in wide binaries from C08a. Our sample has three pairs in common with C08a: 40 Eri A/B, CD-38 10980/CD-30 10983 and L481-60/CD-37 10500. The solid lines in Fig. 6 connect our values to those of C08a. For L481-60/CD-37 10500, both M$\rm_{i}$ and M$\rm_{f}$ are consistent with those in C08a.  However, our M$\rm_{i}$ is different for WDs in the other two pairs. In C08a, the total ages of the binaries were estimated from the X-ray fluxes of their MS components, while our ages are from the $S\rm_{HK}$ of the MS stars. For 40 EriA (HD26965), C08a gives an age of 1.07 Gyr. Our age is 3.56 Gyr, which is closer to its rotation age (4.7 Gyr; Barnes 2007). For CD-38 10983 (HD147513), C08a gives an age of 0.33 Gyr. Our age is 0.55 Gyr, which is also more consistent with the rotation age 0.58 Gyr (Barnes 2007). Moreover, the ages of our 10 pairs are all from the same CA vs. age relation, so no scatter due to the different methods of age determination is imposed.

As can be seen in Table 4, our wide binaries range in age from 0.55 Gyr to 6.54 Gyr. M$\rm_{i}$ ranges from 1.11 M$_{\odot}$ to 4.14 M$_{\odot}$. [Fe/H] ranges from -0.40 to +0.19 dex. Six WDs have M$\rm_{i}$ lower than 2.0 M$_{\odot}$. Thus, wide binaries provide very promising leverage for investigations of the IFMR at masses that are difficult to reach in clusters.

The dotted line in Fig. 6 indicates the semi-empirical IFMR from W00. The dash-dot line is the semi-empirical IFMR from C08b. A theoretical IFMR by Renedo et al. (2010; R10 hereafter) is shown as a dashed line. The empirical IFMR from our wide binary sample is shown as a dash-dot-dot line:

\begin{eqnarray}
\rm{M_{f}} &=& (0.452\pm 0.045) + (0.073\pm0.019) \rm{M_{i}}
\end{eqnarray}

The above relation excludes WD2253-08, the point in Fig. 6 at the upper left. We believe this is an outlier because of its low metallicity, as argued below.

Our IFMR can be adequately represented by a linear function (Equation 5) over the initial mass range 1.1 M$_{\odot}$ to 4.1 M$_{\odot}$. In general, it is similar to other empirical and theoretical relations. However, for higher M$\rm_{i}$, the M$\rm_{f}$ from our IFMR is somewhat lower than other empirical relations. This is probably because only one WD has M$\rm_{i}$ $>$ 3 M$_{\odot}$ in our sample. Like previous IFMR, the scatter in our relation  is also larger than the formal error bars. This suggests other factors may affect the relation. Since the MS component to WD2253-08 has very low metallcity and is the biggest outlier in the relation shown in Fig. 6, we decided to test the influence of metallicity on the IFMR.

Fig. 7 shows lost mass fraction vs. [Fe/H]. Here [Fe/H] is the original metallicity of the WD, as derived from its MS component's spectrum. Two open circles represent DB WDs.  Three triangles are DA WDs whose initial masses were larger than 2 M$_{\odot}$. The other five are DA WDs whose initial masses were lower than 2.0 M$_{\odot}$. The dotted line is a least squares fit for these five DA WDs. There is a clear trend that suggests for a given initial mass, metal rich progenitors lose more mass when they evolve into WDs. We conclude that metallicity plays an important role in the amount of mass lost during post-MS evolution. In addition, the two DBs are both below the dashed line, tentatively suggesting that the metallicity dependence of post-MS mass loss may be different for DB than the DA stars.


\section{Comparison between observations and theoretical models}
Models suggest that the dependence of mass loss rates on stellar parameters is likely steep:  along an evolutionary track R(L, M, Z, mixing length), the slope dlog\.{M}/dlogL from models ranges from $\sim$ 3 for the Wachter et al. (2002) grid carbon-star models to $>$ 16 for some of our model grids.  For such large slopes, the star evolves at essentially constant mass to the death-zone, and leaves the death-zone along an essentially constant core mass track.  The resulting IFMR is very close to what is found by taking the final mass to be the core mass at the death line, M$\rm_{final}$ =  L$\rm_{Death}$(M$\rm_{initial}$).  Note that M$\rm_{initial}$ here is the initial asymptotic giant branch (AGB) mass which may be smaller than the initial MS mass if significant mass loss occurs before the star reaches the AGB tip.  It is known that low-mass, low-Z stars do lose significant mass before they leave the red giant branch (RGB), but we don't know how much mass is lost at that stage by stars with masses significantly greater than 1 M$_{\odot}$.

Our model grids were computed with the Bowen code, using the following parameter values for our standard or ``core" grid: the critical density for onset of density-dependent, less efficient, cooling is 10$^{-12}$ gm/cm$^{3}$; the mean opacity for the atmosphere is $<\rm \kappa_{Rosseland}>$ = 0.0004 cm$^{2}$/gm; the dust condensation temperature is 1350 K with a condensation width $\Delta$ T = 100; and the driving piston at the base of the model is constrained to an amplitude such that the maximum power does not exceed the stellar luminosity.  See Willson (2000) and Willson $\&$ Wang (2011) for details.

In our grids, a model series for fixed M and Z assigned a radius to each L along the AGB track according to the following prescription:

\begin{eqnarray}
\rm R = \rm 312(L/10^{4})^{0.68}(1.175/M)^{0.31S} (Z/0.001)^{0.088}/(l/H)^{0.52}
\end{eqnarray}

where S = 0 for M $<$ 1.175 and 1 for M $>$ 1.175 M$_{\odot}$ and l/H is the ratio of Iben mixing length to scale height (Iben 1984).  We have used the Iben mixing length ratio l/H = 0.9 which gives radii very close to standard evolutionary grids such as the Padova models (Girardi et al. 2000).  A model series would describe the evolution of a single star if it were not for the evolving mass resulting from mass loss; we made model series to study the behavior of the mass loss law.

Metallicity dependence of the mass loss rate can come from two factors:  (a) lower metallicity stars have smaller radii at a given L and M; and (b) lower metallicity stars have a smaller maximum dust/gas ratio which helps expel material.  For dlogR/dlogZ = 0.088 (Iben 1984) and dlog\.{M}/dlogR $\sim$ 10 (our models) we expect dlog\.{M}/dlogZ $\sim$ 0.9. This translates into dlogL$\rm_{death}$/dlogZ $\sim$ 0.1 for typical values of dlog\.{M}/dlogR and dlog\.{M}/dlogL.  The final mass is approximately M$\rm_{core}$(L$\rm_{Death}$(M$\rm_{iAGB}$)) allowing us to link M$\rm_{i}$ to M$\rm_{f}$ via L$\rm_{death}$.

Fig. 8 shows the agreement between our observations and two model grids. Model A (left panel) is our ``core" case Bowen model described above, with no mass loss prior to the AGB.  Varying model parameters (critical density for cooling, opacity, dust condensation temperature, or mixing length) has a small effect on the position of the $\Delta$M/M$\rm_{i}$ vs Z points and almost no effect on the slope. However, it is known that at least some stars lose mass on the RGB.  Model B (right panel) assumes that only low-mass stars lose appreciable mass on the RGB:  $\Delta$M$\rm_{RGB}$ = (2 - M$\rm_{i}$)*0.15 for 1 M$_{\odot}$ $<$ M$\rm_{i}$ $<$ 2 M$_{\odot}$. We computed the standard deviation of the residuals, $\sqrt{\sum(\Delta\rm M_{observed}-\Delta M)^{2}/5}$ = SD5 for the 5 stars and the standard deviation for the four stars with the smallest error bars (SD4). 
Including RGB mass loss by the above prescription improves the fit slightly for the 5-star comparison and substantially for the 4-star comparison.

As stars evolve up the AGB, some have their envelopes enriched in carbon.  As C/O rises, the radius for a given L, M, Z increases, and this will in turn enhance the mass loss rate at a given L (or core mass) and decrease the deathline L for a given M, Z.  Using Figure 19 of Marigo $\&$ Girardi (2007) as a guide, we find that of the stars in our sample, only the one with initial mass = 2.9 solar masses is likely to have been a carbon star.  However, the star with the lowest metallicity is close to the line dividing stars that do from stars that don't, and any enhancement, even if C/O remains $<$ 1, will increase the radius and decrease the deathline L.

Overall, the mass loss models predict net mass loss and M$\rm_{f}$ vs. M$\rm_{i}$ that are very close to the values found from the WD+MS pairs. This will be true for other mass loss formulae with similar death-lines, as discussed by Willson (2007, 2008, 2009) and Willson et al. (2008). In addition, the Z-dependence of our models is consistent with the data from these WD+MS pairs. It is important to note that the bulk of the Z-dependence in the mass loss rates comes from the dependence of R on Z, with a smaller effect from the efficiency of forming dust (the gas/dust ratio) in the models.  Also, if C/O is increased over solar values this will decrease the final mass for a given initial mass and metallicity.


\section{Conclusion}
In this study, we constructed an empirical IFMR using 10 WDs in wide binaries.
Our IFMR contains six WDs whose M$\rm_{i}$ are lower than 2 M$_{\odot}$. They contribute to the low initial mass limit that is not well-sampled by clusters.  Our WDs in wide binaries suggest a linear IFMR over the initial mass range 1.1 M$_{\odot}$ to 4.1 M$_{\odot}$ (Equation 5).

We compared our mass loss vs. metallicity relation to theoretical models for evolving lower MS stars ($<$ 2 M$_{\odot}$). In general, the models predict a net mass loss and IFMR that agree with the values found from our observation within the current uncertainty of measurement.

Kalirai et al. (2007) and Kalirai et al. (2009) tentatively found a metallicity dependence on the IFMR.
We find that at least part of the scatter seen in the IFMR is correlated with metalllcity. Stars with lower metallicity apparently shed less mass when they become WDs.

\acknowledgments
Many thanks to D. Koester for providing his WD models. Balmer/Lyman lines in the models were calculated with the modified
Stark broadening profiles of Tremblay $\&$ Bergeron, ApJ 696, 1755, 2009,
kindly made available by the authors. T.D.O. acknowledges support from NSF grant AST-0807919
to Florida Institute of Technology. J.K.Z. and G.Z. acknowledge support from NSFC grant Nos. 11078019 and 10821061. L.A.W. and Q.W. acknowledge support from NSF grant AST-0708143.

\clearpage




\clearpage

\clearpage
\begin{figure}
\epsscale{0.8}
\plottwo{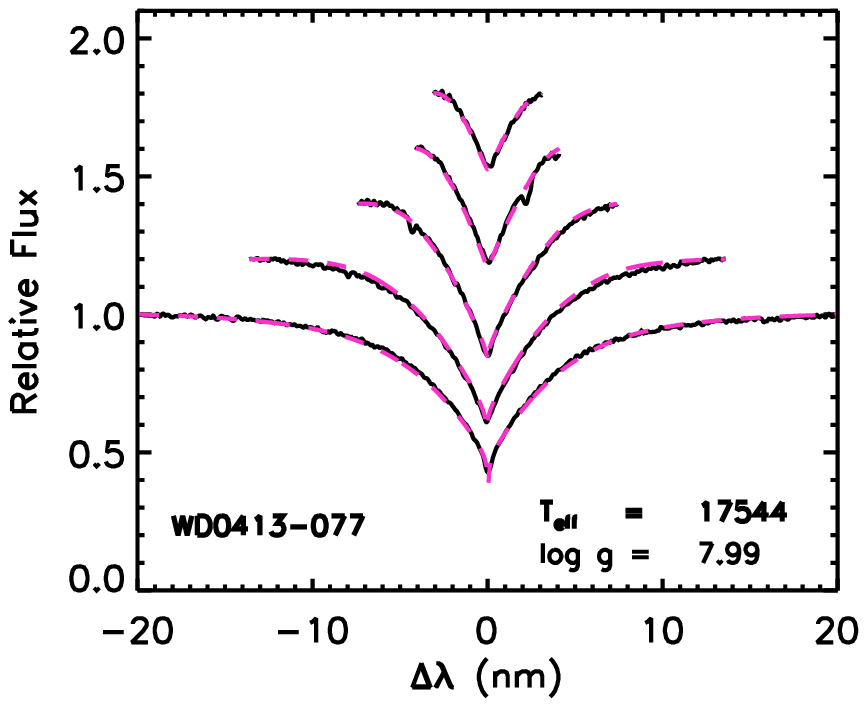}{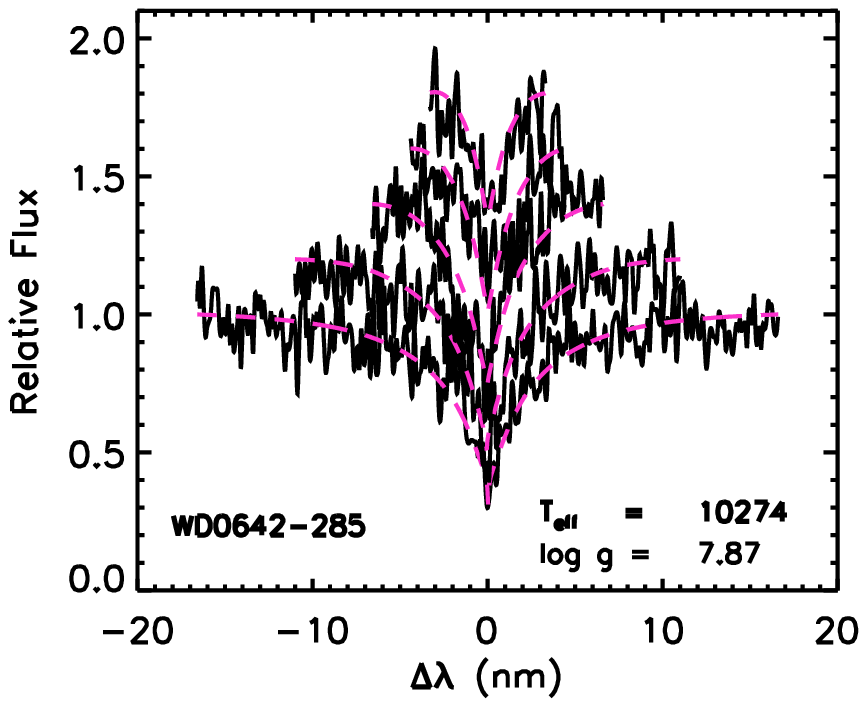}
\plottwo{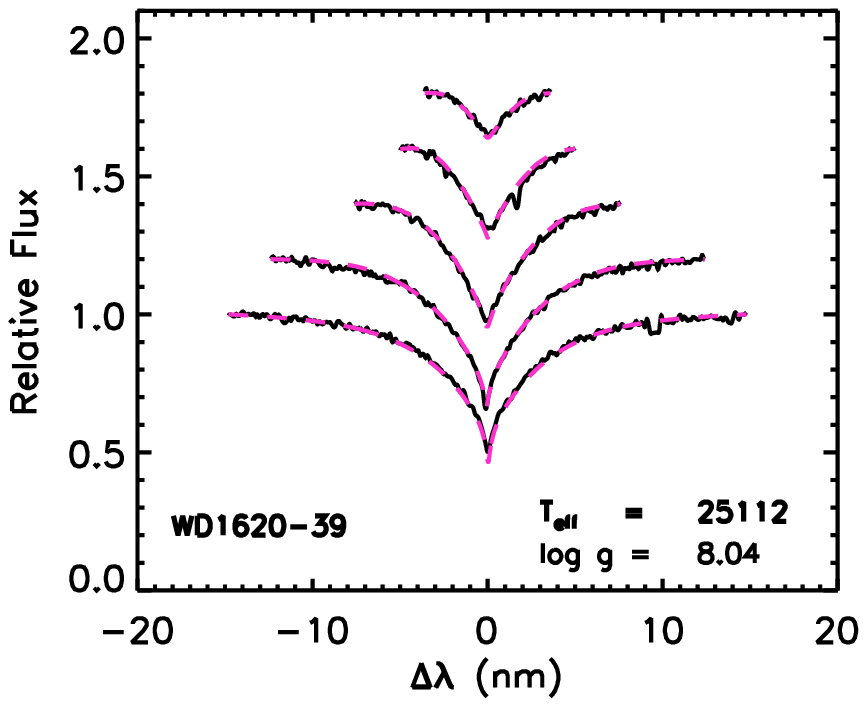}{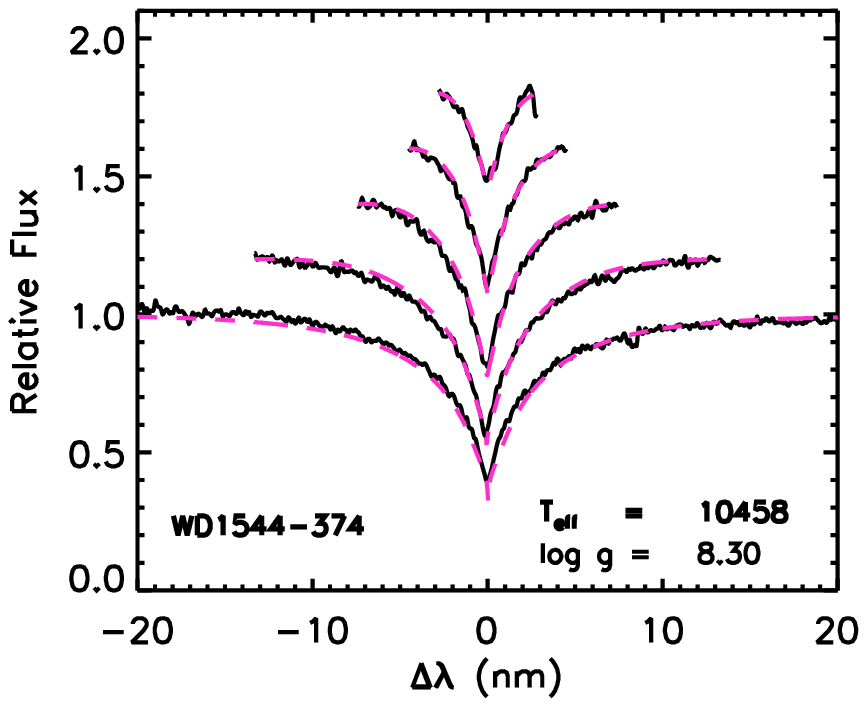}
\epsscale{0.8}
\plottwo{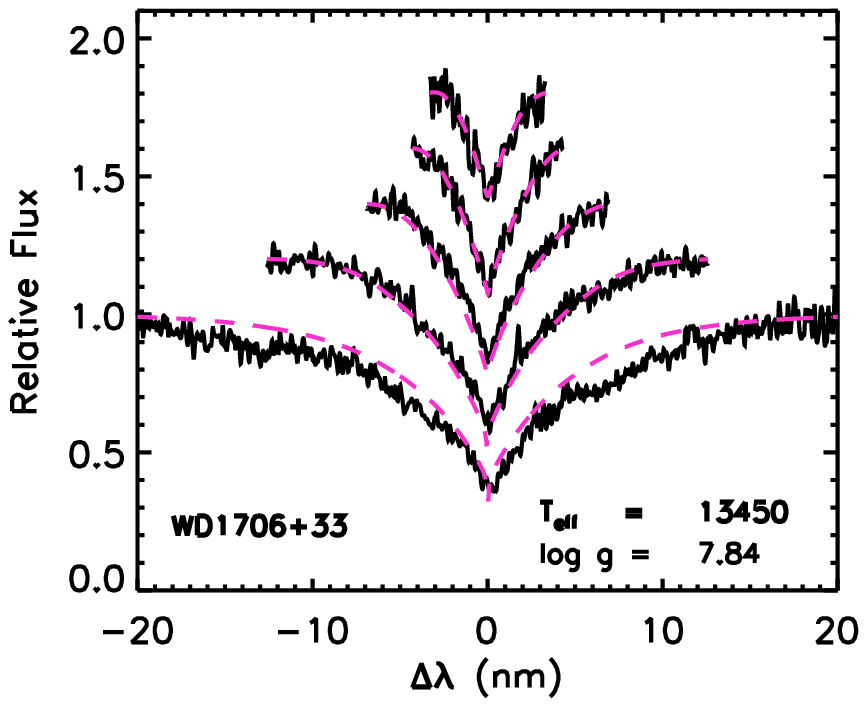}{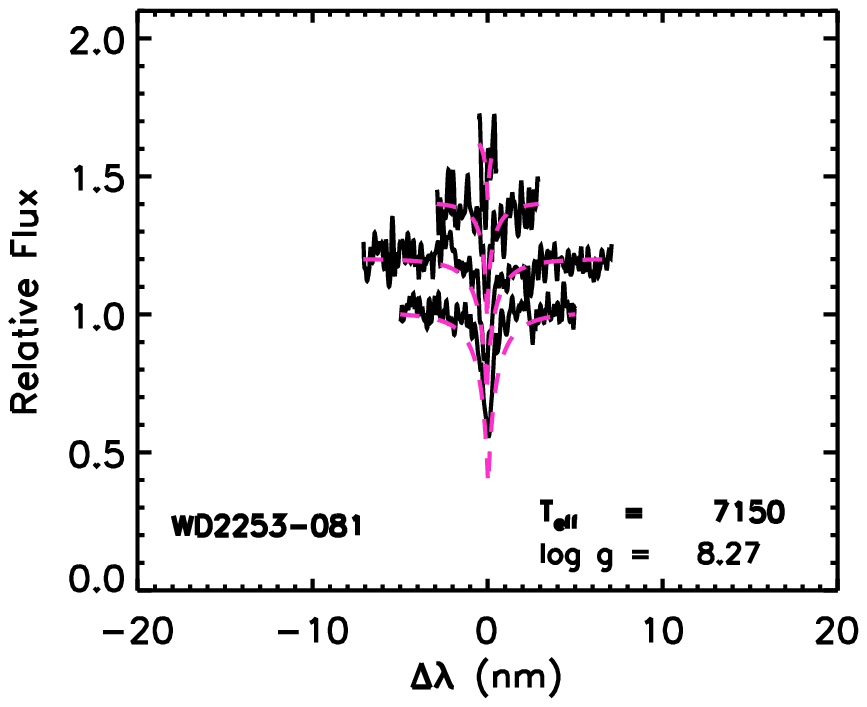}
\plottwo{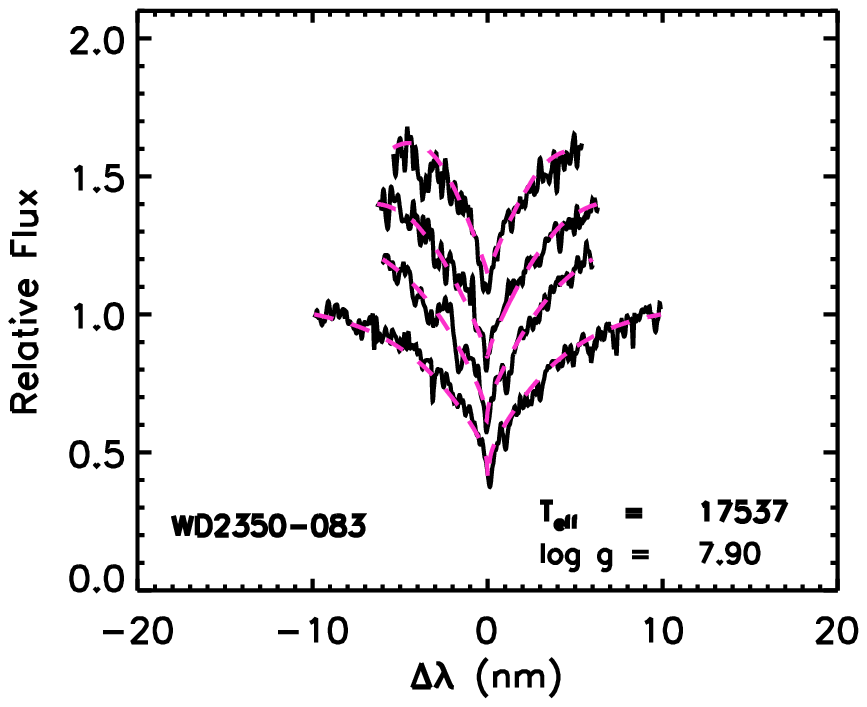}{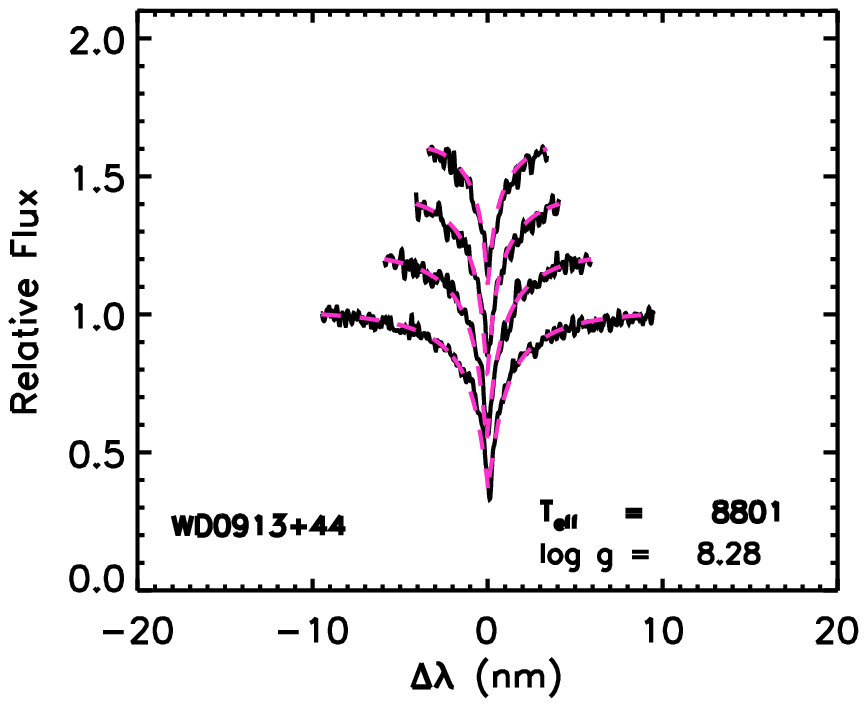}
\epsscale{0.4}
\plotone{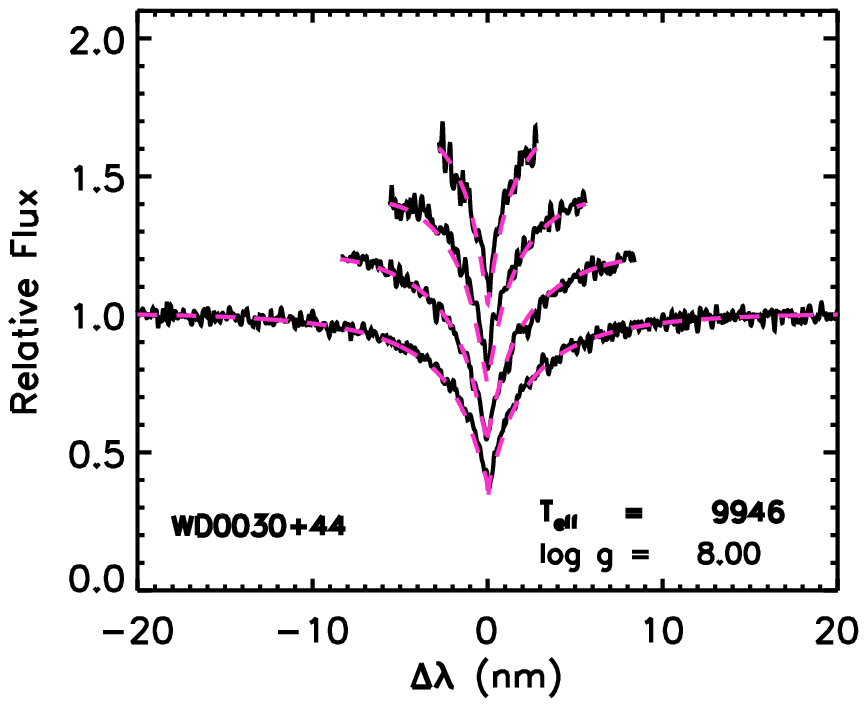}{}

\caption{Fits of the observed Balmer lines for the white dwarfs studied here. Lines range from H$\beta$ (bottom) to H8 (top) or H$\delta$}
\end{figure}

\clearpage
\begin{figure}
\epsscale{1.0}
\plotone{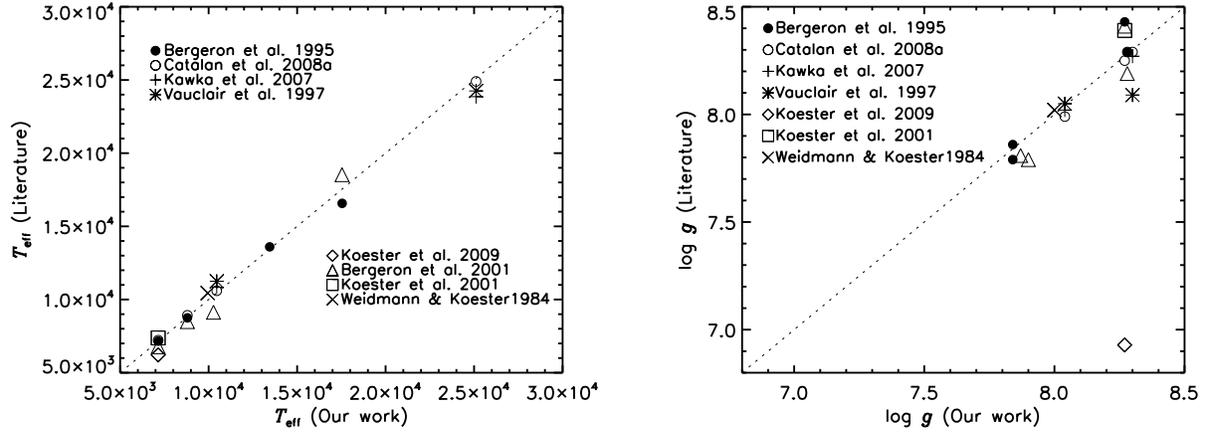}{}
\caption{$T\rm_{eff}$ and log $g$ comparison between our results and those from the literature. The dotted line is the unit slope relation. }
\end{figure}

\clearpage
\begin{figure}
\epsscale{.75}
\plotone{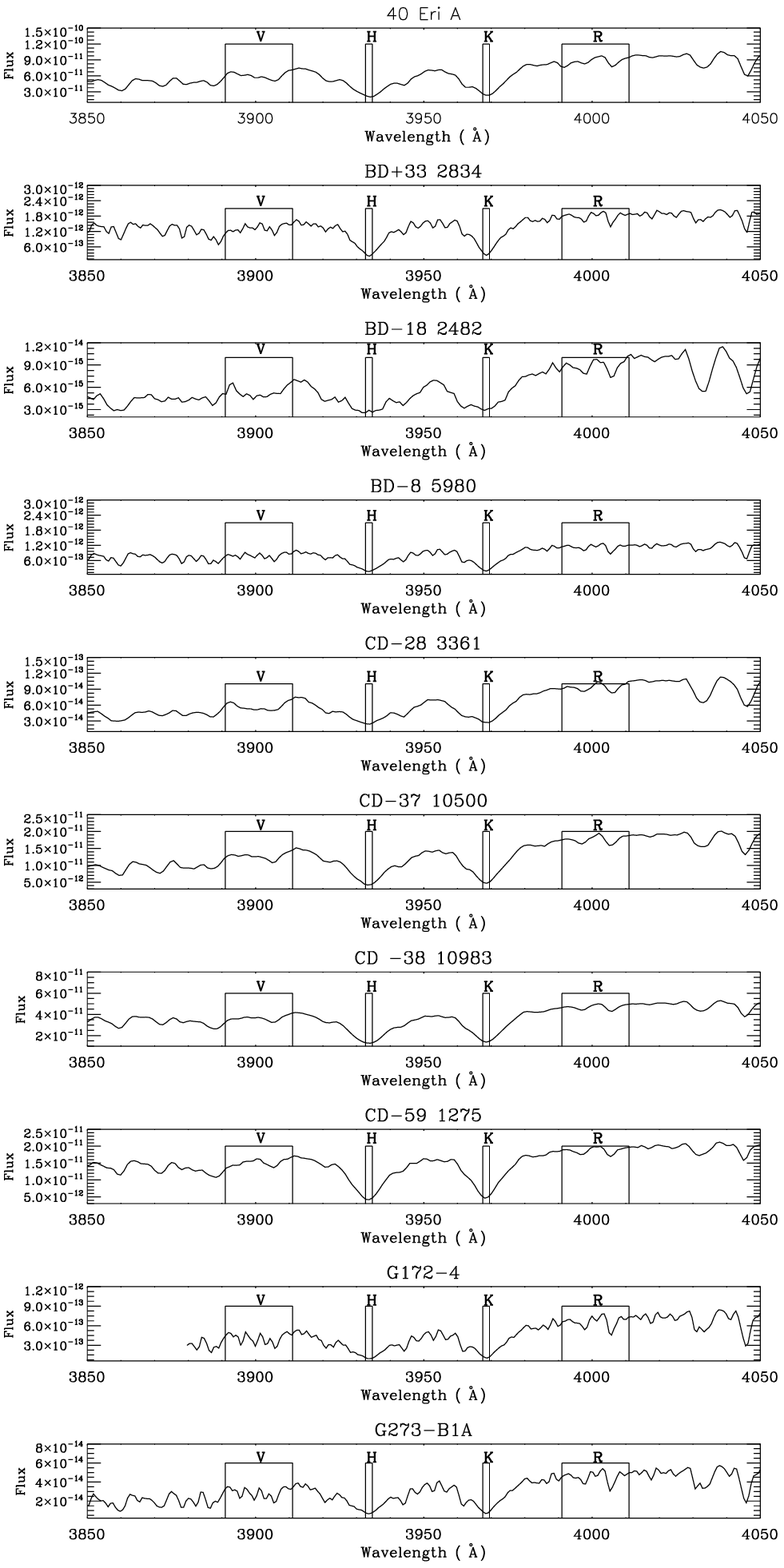}{}
\caption{The $S$$\rm_{HK}$ measurements in 10 MS stars. }
\end{figure}

\clearpage
\begin{figure}
\epsscale{0.8}
\plotone{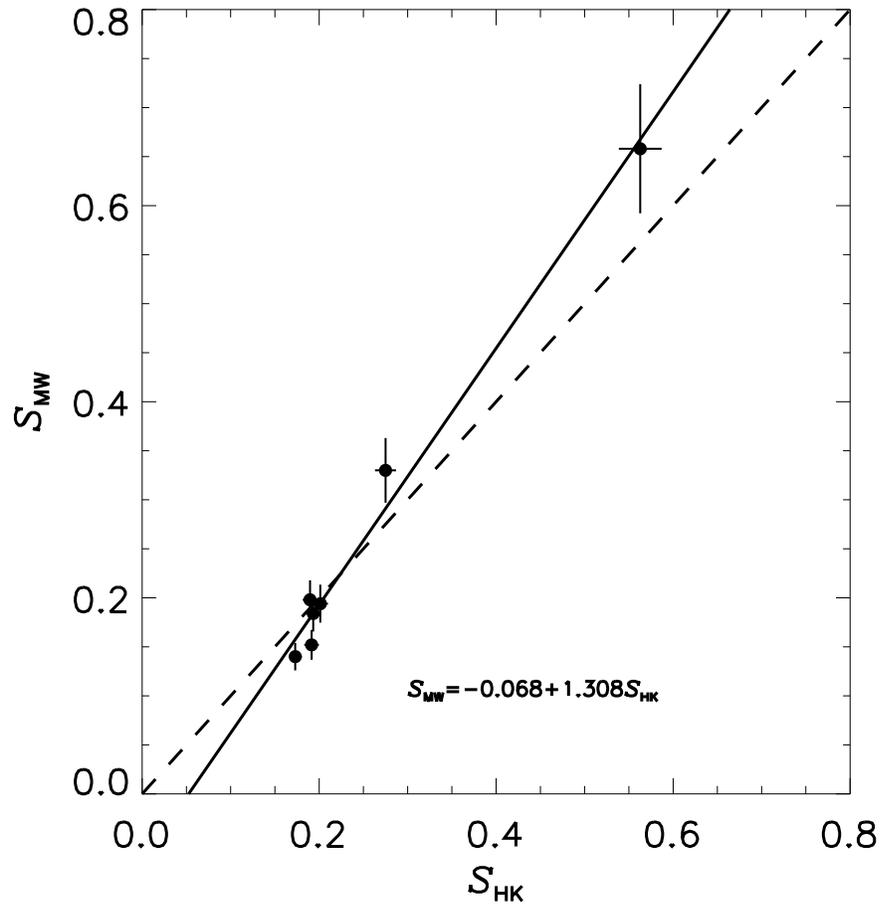}{}
\caption{Calibration between our chromospheric
activity index $S\rm_{HK}$ and the Mount Wilson $S\rm_{MW}$ index. The solid line is the least squares fit, while a dashed line is the unit slope relation.}
\end{figure}

\clearpage
\begin{figure}
\epsscale{1.0}
\plotone{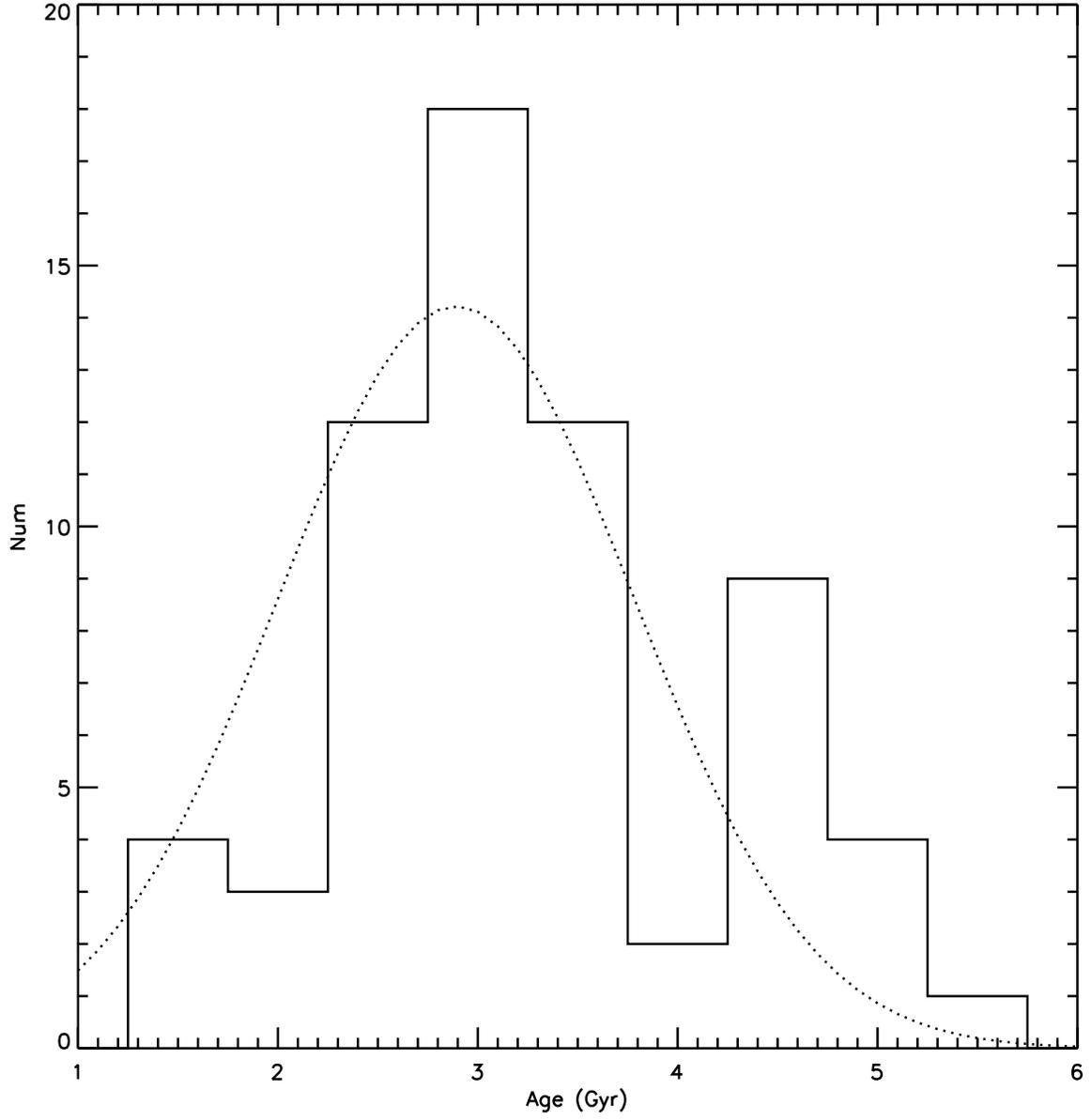}{}
\caption{The chromospheric activity distribution of M67 member stars. The dotted line is the Gaussian fitting.}
\end{figure}

\clearpage
\begin{figure}
\epsscale{1.0}
\plotone{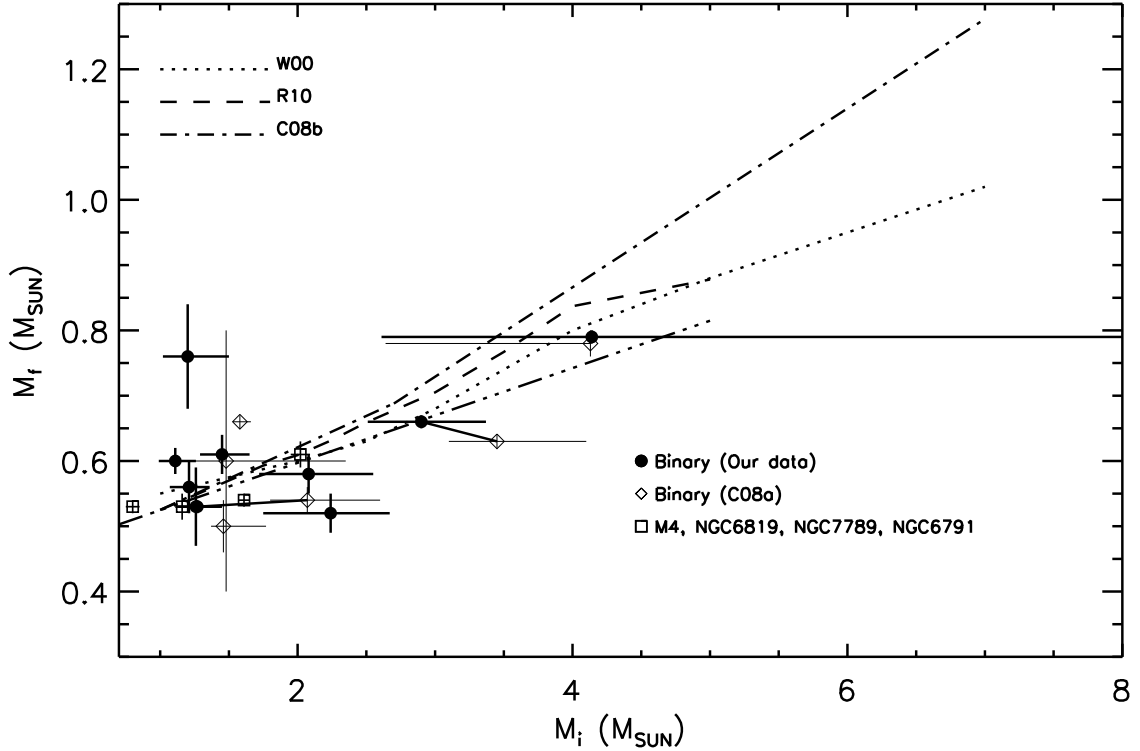}{}
\caption{ WD initial-final mass relationship shown for the 10 WDs
in this work (filled circles). The Diamonds represent the WDs in wide binaries from Catal\'{a}n et al. (2008a). The dotted line is the empirical IFMR of Weidemann (2000). The dashed line is the theoretical IFMR (Z = 0.01) from Renedo et al. (2010). The dashed dot line is the empirical IFMR from Catal\'{a}n et al. (2008b). The dash-dot-dot line is a least squares fit of 9 WDs (not including the left top star WD2253-08) in our paper. The solid lines connect the WD points both in our paper and Catal\'{a}n et al. (2008a). The squares represent mean M$\rm_{f}$ and M$\rm_{i}$ values of WDs in four clusters: M4, NGC6819, NGC7789 and NGC6791. }
\end{figure}

\clearpage
\begin{figure}
\epsscale{1.0}
\plotone{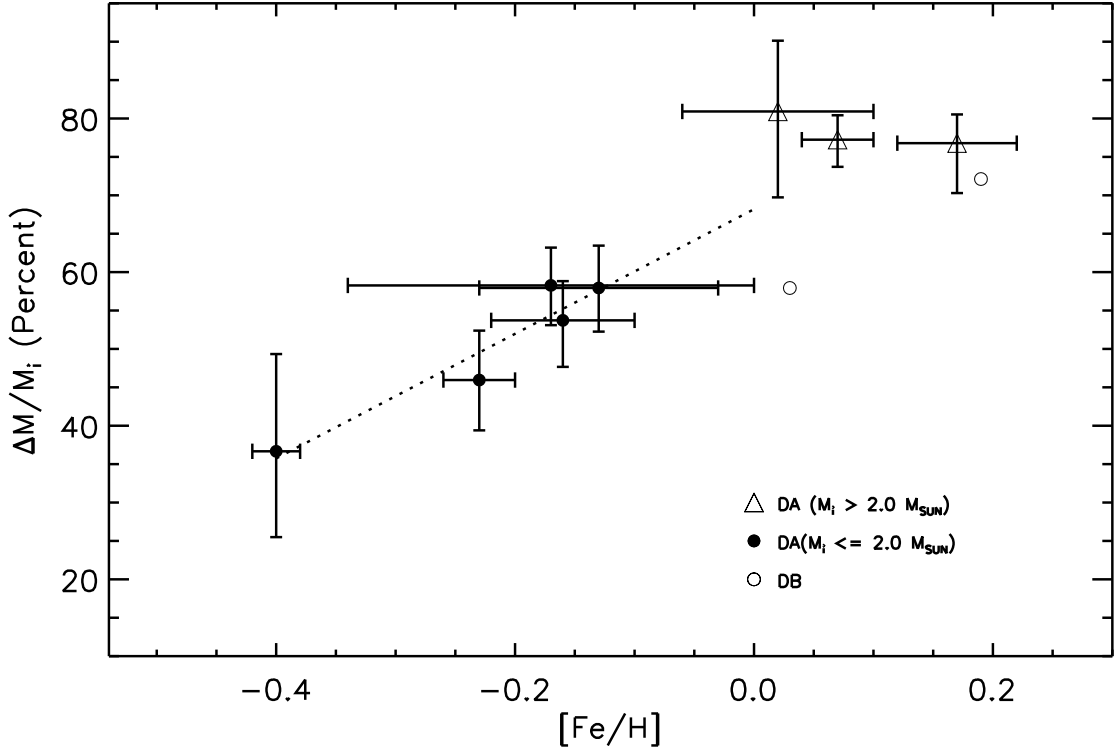}{}
\caption{Lost mass fraction vs. [Fe/H]. Triangles are WDs with M$\rm_{i}$ $>$ 2.0 M$_{\odot}$.  Filled circles are DA WDs and open circles are DB WDs. Dotted line is the fit from 5 DA WDs with M$\rm_{i}$ smaller than 2.0 M$_{\odot}$. }

\end{figure}

\begin{figure}
\epsscale{1.0}
\plottwo{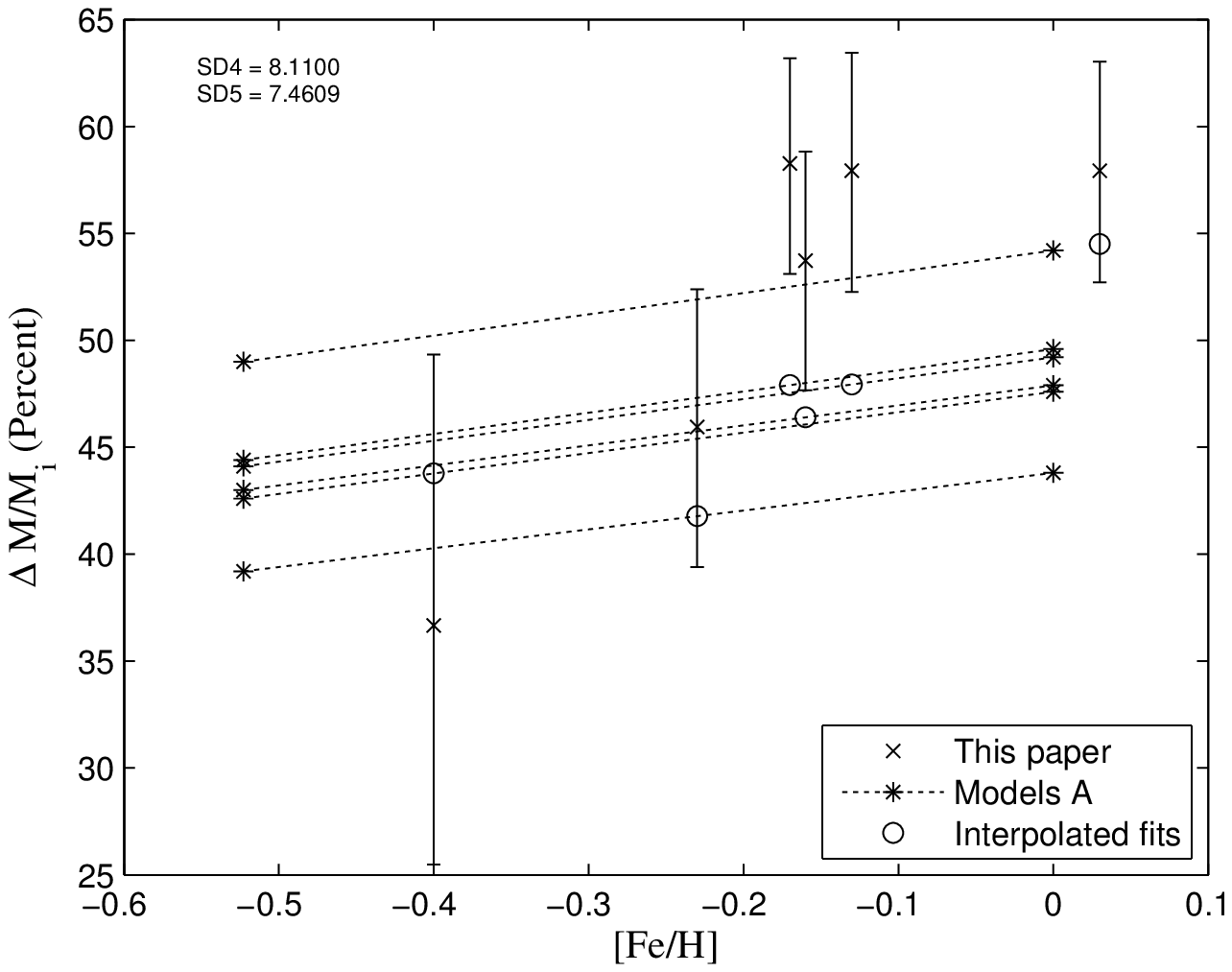}{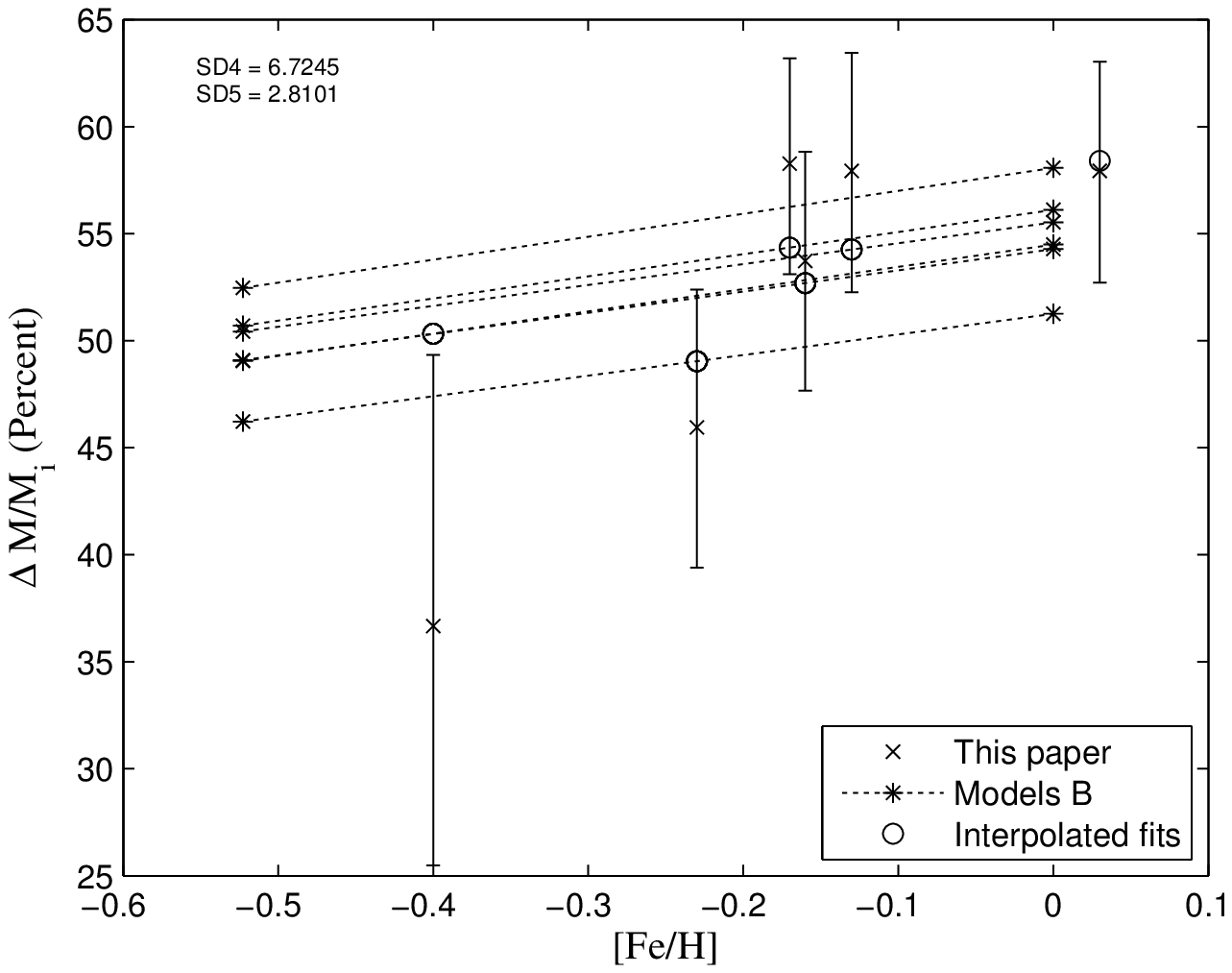}
\caption{Comparison of observed lost mass fraction from Figure 6 with two model sets. In each graph, the points marked by asterisks are derived as described in the text from a series of models with the same masses as the observed stars but with two different values for Z. The fits for each of the observed stars (open circles) are then obtained by interpolating between the two Z model sets along a fixed-mass (dashed) line. Case A is our standard ``core" model as described in the text. To test the effects of mass loss by low mass stars ($<$ 2 M$_{\odot}$) along the RGB or at the core flash, we also computed case B with a prescription for RGB mass loss given by $\Delta$M$\rm_{RGB}$ = 0.15 (2-M$\rm_{initial}$) for 1 $<$ M$\rm_{initial}$ $<$ 2. This was chosen as the simplest rule giving reasonable $\Delta$M$\rm_{RGB}$ for M = 1 and M = 2. SD5 and SD4 are the standard deviations of the residuals including 5 and 4 stars (the latter omitted the one with the largest error bars).}  
\end{figure}


\begin{deluxetable}{lcclccc}
\tabletypesize{\scriptsize}
\tablecaption{Wide binaries in our sample \label{tbl-1}}
\tablewidth{0pt}
\tablehead{
\colhead{Wide Binary}& \colhead{RA}& \colhead{Dec}&\colhead{White Dwarf}&\colhead{Spectral Type} & \colhead{UT}&\colhead{Site}\\

\colhead{(1)}&(2)&(3)&\colhead{(4)}&(5)&(6)&(7)\\
}
\startdata
40 Eri A/B& 04 13 03& -07 44 06&WD0413-077&DA&02/04&CTIO \\
CD-59 1275/L182 61& 06 15 36& -59 11 24&WD0615-591&DB&02/04&CTIO \\
CD-28 3361/LP895-41& 06 42 34& -28 30 48&WD0642-28&DA&02/04&CTIO \\
BD-18 2482/LP786-6& 08 45 18 &-18 48 00&WD0845-188&DB&02/04&CTIO \\
CD-38 10983/10980&16 20 38 &-39 04 42 &WD1620-39&DA&02/04&CTIO \\
CD-37 10500/L481-60&15 44 12& -37 46 00 &WD1544-374&DA&02/04&CTIO \\
BD+33 2834/G181-B5B&17 06 58&	33 16 48&WD1706+33&DA&07/10&KPNO \\
BD-8 5980/G156-64&22 53 12&	-08 05 24&WD2253-08&DA&07/10&KPNO \\
G273-B1A/B&23 50 54&-08 21 06&WD2350-083&DA&11/06&KPNO\\
BD+44 1847/G116-16\tablenotemark{a}&09 11 51&44 15 36&WD0913+44&DA&11/06&KPNO\\
G172-4/G171-62&00 30 17 &44 27 18 &WD0030+44&DA&11/06&KPNO\\

\enddata
\tablenotetext{a}{Non-physical pair; see text.}

\end{deluxetable}

\begin{deluxetable}{ccccccccc}
\tabletypesize{\scriptsize}
\tablecaption{Physical properties of WDs in wide binaries \label{tbl-1}}
\tablewidth{0pt}
\tablehead{
\colhead{Name}&S/N\tablenotemark{a}&\colhead{$T$$\rm_{eff}$}& \colhead{log $g$}& \colhead{M$\rm_{f}$} & \colhead{t$\rm_{cool}$}&\colhead{t$\rm_{progenitor}$}&\colhead{[Fe/H]}&\colhead{M$\rm_{i}$}\\
&(K)&&(M$_{\odot}$)&(Gyr)&(Gyr)&&(M$_{\odot}$)\\
(1)&(2)&(3)&(4)&(5)&(6)&(7)&(8)\\
}
\startdata
WD0413-077 &240& 17544$\pm$64 &7.84$\pm$0.01 &0.53$\pm$0.01  &0.086$\pm$0.028&3.48$_{2.33}^{5.19}$&-0.17$\pm$0.17&1.27$_{1.13}^{1.44}$ \\
WD0642-28 &57& 10274$\pm$122 &7.87$\pm$0.12 &0.53$\pm$0.06  &0.482$\pm$0.077&3.67$_{2.32}^{5.66}$&-0.13$\pm$0.10&1.26$_{1.11}^{1.45}$ \\
WD1620-39 &156& 25112$\pm$102 &8.04$\pm$0.01 &0.66$\pm$0.01  &0.025$\pm$0.001&0.52$_{0.35}^{0.79}$&0.07$\pm$0.03&2.90$_{3.37}^{2.51}$ \\
WD1544-374 &242& 10458$\pm$25 &8.30$\pm$0.02 &0.79$\pm$0.01  &0.844$\pm$0.025& 0.20$_{0.01}^{0.70}$&0.02$\pm$0.08&4.14$_{2.61}^{8.00}$\\
WD1706+33 &34& 13450$\pm$453 &7.84$\pm$0.05 & 0.52$\pm$0.03  &0.219$\pm$0.038&1.08$_{0.66}^{1.71}$&0.17$\pm$0.05&2.24$_{1.75}^{2.67}$ \\
WD2253-08 &45& 7150$\pm$67 &8.27$\pm$0.12 &  0.76$\pm$0.08  &2.470$\pm$0.739&4.07$_{1.95}^{7.20}$&-0.40$\pm$0.02&1.20$_{1.02}^{1.50}$ \\
WD2350-083 &52& 17537$\pm$266 &7.90$\pm$0.07 &  0.56$\pm$0.04  &0.097$\pm$0.015&4.18$_{2.79}^{6.28}$&-0.16$\pm$0.06&1.21$_{1.07}^{1.36}$ \\
WD0030+44 &46& 9946$\pm$34 &8.00$\pm$0.04 &  0.60$\pm$0.02  &0.612$\pm$0.041&5.38$_{3.44}^{8.25}$&-0.23$\pm$0.03&1.11$_{0.99}^{1.26}$ \\
WD0615-591\tablenotemark{b} & $<$20 & 16714$\pm$200 &8.02$\pm$0.05 &  0.61$\pm$0.03  &0.155$\pm$0.016&3.01$_{1.99}^{4.53}$&0.03$\pm$0.07&1.45$_{1.29}^{1.65}$ \\
WD0845-188\tablenotemark{b} &$<$20 & 17566$\pm$350 &7.97$\pm$0.05 &   0.58$\pm$0.03  &0.118$\pm$0.013&1.18$_{0.76}^{1.80}$&0.19$\pm$0.09&2.08$_{1.72}^{2.55}$ \\
\enddata
\tablenotetext{a}{Average S/N surrounding H$\delta$.}
\tablenotetext{b}{The type of WD is DB. $T$$\rm_{eff}$ and log $g$ are from Voss et al. (2007).  }
\end{deluxetable}

\begin{deluxetable}{cccc}
\tabletypesize{\scriptsize}
\tablecaption{CA standard stars}
\tablewidth{0pt}
\tablehead{
\colhead{Name}&\colhead{B-V}& \colhead{$S\rm_{HK}$}& \colhead{$S\rm_{MW}$}\\
(1)&(2)&(3)&(4)\\}
\startdata
HD212754 & 0.520 & 0.173&0.140 \\
HD206860 & 0.590 &0.275 &0.330 \\
HD207978& 0.420 &0.192 &0.152 \\
HD224930 & 0.670 &0.193 &0.184 \\
HD10476 & 0.840 &0.190 &0.198 \\
HD190406&0.610&0.201&0.194\\
\enddata
\end{deluxetable}

\begin{deluxetable}{lccccc}
\tabletypesize{\scriptsize}
\tablecaption{Ages of the wide binaries estimated by CA }
\tablewidth{0pt}
\tablehead{
\colhead{Name}&\colhead{B-V}&\colhead{$S\rm_{HK}$}&\colhead{$S\rm_{MW}$}& \colhead{$R\arcmin\rm_{HK}$}& \colhead{age}\\
\colhead{(1)}&(2)&(3)&(4)&(5)&(6)\\}
\startdata
40 Eri A/B & 0.775& 0.305&0.206&-4.85&3.56$_{2.41}^{5.28}$ \\
CD-59 1275/L182-61 & 0.590 &0.268 &0.156& -4.99&3.17$_{2.14}^{4.68}$\\
CD-28 3361/LP895-41 & 0.972 &0.346 &0.238&-4.92&4.15$_{2.81}^{6.14}$\\
BD-18 2482/LP786 & 1.036 &0.407 &0.307& -4.88&1.29$_{0.87}^{1.91}$\\
CD-38 10983/CD-38 10980 &0.630 &0.325 &0.29& -4.52&0.55$_{0.37}^{0.81}$ \\
CD-37 10500/L481-60 & 0.718 &0.303 &0.256& -4.66&1.05$_{0.71}^{1.55}$ \\
BD+33 2834/G181-B5B & 0.568 &0.188 &0.171& -4.87&1.30$_{0.88}^{1.93}$\\
BD-8 5980/G156-64 & 0.630 &0.197 &0.188& -4.84&6.54$_{4.42}^{9.67}$ \\
G273-B1A/B & 0.770 &0.196 &0.188& -4.91&4.28$_{2.89}^{6.32}$\\
G171-62/G172-4 & 0.980 &0.230 &0.232& -4.94&5.99$_{4.05}^{8.86}$\\
\enddata
\end{deluxetable}

\begin{deluxetable}{ccccc}
\tabletypesize{\scriptsize}
\tablecaption{Comparison between our ages and those from the literature \label{tbl-2}}
\tablewidth{0pt}
\tablehead{
\colhead{Name} &\colhead{age\tablenotemark{a}}& \colhead{age\tablenotemark{b}}& \colhead{rotation age\tablenotemark{c}}& \colhead{our age} \\
&(Gyr)&(Gyr)&(Gyr)&(Gyr)\\
(1)&(2)&(3)&(4)&(5) \\
  }
\startdata
 40 Eri A&$\sim$&12.2$_{8.5}^{14.5}$&4.75$\pm$0.75&3.56$_{2.41}^{5.28}$ \\
 CD-38 10983&2.5$^{7.0}_{\sim}$&2.0$_{0.4}^{3.9}$&0.58$\pm$0.08&0.55$_{0.37}^{0.81}$ \\
 CD-59 1275&5.9$_{5.4}^{6.6}$&3.7$_{3.4}^{4.7}$&&3.17$_{2.14}^{4.68}$ \\
 CD-37 10500&7.4$_{1.9}^{13.0}$&4.4$_{1.4}^{7.0}$&&1.05$_{0.71}^{1.55}$ \\
 \enddata
\tablenotetext{a}{ages are from Holmberg, Nordstr\"{o}m $\&$ Andersen (2009) }
\tablenotetext{b}{ages are from Valenti $\&$ Fischer (2005) }
\tablenotetext{c}{ages are from Barnes (2007) }

\end{deluxetable}


\clearpage




\end{document}